\documentclass{PoS} % BB Sep 03 2011.

\title{Finite Volume Corrections to the SU(3) Deconfining Temperature 
due to a Confined Exterior}

\ShortTitle{Finite Volume Corrections to the SU(3) Deconfining 
            Temperature}

\author{\speaker{Bernd A. Berg} and Hao Wu\\ Department of Physics,
        Florida State University, Tallahassee, FL 32306, USA}
\abstract{Deconfined regions in relativistic heavy ion collisions are 
limited to small volumes surrounded by a confined exterior. Using the 
geometry of a double layered torus, we keep an outside temperature 
slightly lower than the inside temperature, so that both regions are 
in the SU(3) scaling region. Deconfined volume sizes are chosen to be 
in a range typical for such volumes created at the BNL RHIC. Even with 
small temperature differences a dependence of the (pseudo) deconfining 
temperature on a colder surrounding temperature is clearly visible. For 
temporal lattice sizes $N_{\tau}=4$, 6 and 8 we find consistency with 
SU(3) scaling behavior for the measured transition temperature signals.} 

\FullConference{The XXIX International Symposium on Lattice Field Theory\\
                July 11-16, 2011\\ Squaw Valley, CA, USA}

\begin{document}

\section{Introduction \label{sec:intro} }

In relativistic heavy ion collisions (RHIC) we are dealing with an 
ensemble of volumes with typical length scales of 5 to 10 fermi.
Particular volumes depend on details of the collision. These length 
scales are not really small when compared with a characteristic 
correlation length set by the inverse (infinite volume) deconfining
temperature $T_c$. Estimates of $T_c$ vary by up to 20\% \cite{Ludmila}, 
but for our SU(3) investigation an approximate value is entirely 
sufficient, which we choose as in earlier work \cite{BB07} to be
\begin{equation}
  T_c = 174\,{\rm MeV}
\end{equation}
and hence $T_c^{-1} = 1.13\,{\rm fermi}$, 10\% to 20\% of a typical 
volume extension. Therefore, one can expect finite volume corrections
due to cold boundaries, while with some notable exceptions 
\cite{BB07,FV} most work in the literature addressed the infinite 
volume limit.

In lattice simulations of pure SU(3) gauge theory equilibrium 
configurations at temperature
\begin{equation} \label{eq:T}
  T = \frac{1}{a(\beta^g)\,N_{\tau}}
\end{equation}
can be generated by Markov chain Monte Carlo (MCMC) on 4D hypercubic
lattices of size $N_{\tau}\,N_s^3$, $N_{\tau}<N_s$ with lattice 
spacing $a$ and $\beta^g=6/g^2$, where $g$ is the bare SU(3) coupling. 
The infinite volume limit is quickly approached through use of 
periodic boundary conditions (PBCs). Obviously, they are not 
suitable for our purposes and need to be replaced by boundary 
conditions (BCs) which model the confined phase.

In~\cite{BB07} zero outside temperature defined by the strong 
coupling limit $a(\beta^g)\to\infty$ for $\beta^g\to 0$, called cold 
boundary condition (CBC), was targeted. 
For realistically sized volumes corrections in the range 
from 30~MeV down to 10~MeV were found compared to signals of the 
deconfinement transition relevant for large volumes. Although 
agreement with SU(3) scaling was seen when varying $N_{\tau}$ from 
4 to~6 and~8, one may be worried because the construction mixes a 
SU(3) scaling region with BCs from the extreme strong coupling limit. 

\begin{figure} \begin{center}
\includegraphics[width=0.7\textwidth]{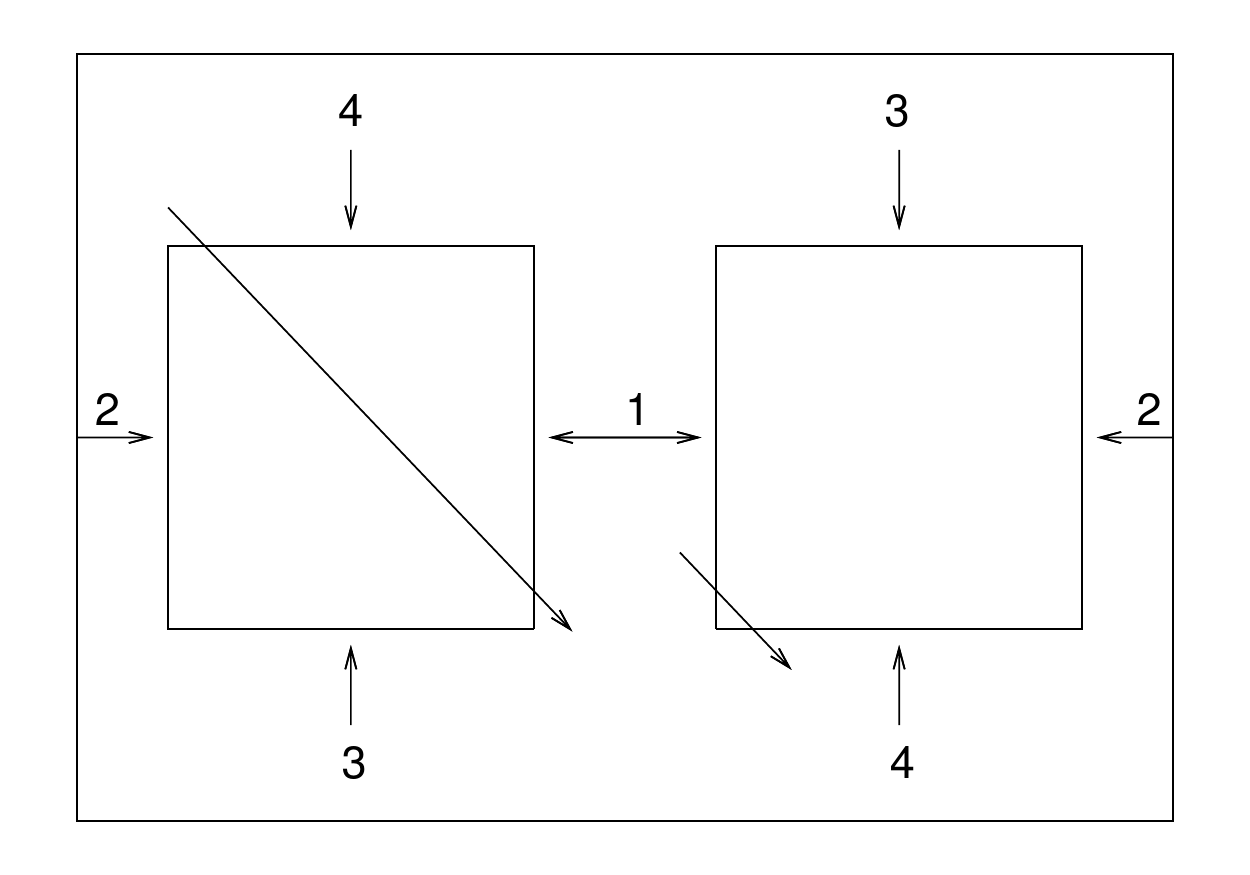} 
\caption{Double layered torus in two dimensions.} \label{fig_DLT}
\end{center} \end{figure} 

To use BCs from SU(3) configurations, which are also in the SU(3) 
scaling regions, the construction of a double-layered torus (DLT) 
was explored in \cite{BW09,BBW09}. For the DLT boundaries are glued
together as indicated for the 2D case by the arrows in 
Fig.~\ref{fig_DLT} (the 2D topology is that of a Klein bottle). In 
the next section we discuss preliminaries for our investigation and 
in section~\ref{sec:prelim} we report simulations using DLT BCs in 
3D and PBCs in $N_{\tau}$ direction. We keep one DLT layer, called 
``outside'', below $T_c$ at $T_{\rm out}\approx 160\,{\rm MeV}$ and 
search for signals of the deconfining transition in the other layer
by varying its temperature.

\section{Preliminaries \label{sec:prelim} }

We use the single plaquette {\it Wilson action} on a 4D hypercubic 
lattice. Numerical evidence \cite{A01} supports that SU(3) lattice 
gauge theory exhibits a weakly first-order deconfining phase transition 
at coupling constant values $\beta^g_c(N_{\tau})=6/g^2_c(N_{\tau})$. 
{\it The scaling behavior} of the deconfining temperature is
\begin{equation} 
  T^c = c_T\,\Lambda_L
\end{equation}
where the lambda lattice scale
\begin{equation} \label{eq:LambdaL}
  a\,\Lambda_L = f_{\lambda}(\beta^g) = \lambda(g^2)\,
  \left(b_0\,g^2\right)^{-b_1/(2b_0^2)}\,e^{-1/(2b_0\,g^2)}\,,
\end{equation}
has been {\it determined in the literature}. The coefficients $b_0$ 
and $b_1$ are perturbatively obtained from the renormalization group 
equation,
\begin{equation}
  b_0 = \frac{11}{3}\frac{3}{16\pi^2}~~{\rm and}~~
  b_1=\frac{34}{3}\left(\frac{3}{16\pi^2}\right)^2\,.
\end{equation}
Using results from \cite{Boyd96}, higher perturbative and 
non-perturbative corrections are parametrized 
\cite{BBV06} by
\begin{equation} 
  \lambda(g^2)\ =\ 1+e^{\ln a_1}\,e^{-a_2/g^2}+a_3\,g^2+a_4\,g^4
\end{equation}
with $\ln a_1=18.08596,\ a_2=19.48099,\ a_3=-0.03772473,\ a_4=0.5089052.$
In the region accessible by MCMC simulations this parametrization is 
perfectly consistent with an independent earlier one \cite{NS02} (less 
than 1\% deviation in the range of validity of \cite{NS02}) and has 
the advantage to reduce for $g^2\to 0$ to the perturbative limit.

For PBC finite size corrections to the deconfining temperature $T_c$
are negligible, so that we can use the $N_{\tau}=4,\,6,\,8$ estimates
of the transition coupling $\beta^g_c(N_{\tau})$ from \cite{Boyd96}
to fix $\Lambda_L$. Using (\ref{eq:T})
\begin{equation} 
  \Lambda_L = a^{-1}\,f_{\lambda}(\beta^g_c) =
  a^{-1}\,f_{\lambda}(\beta^g_c)\ a\,N_{\tau}\,T_c = 
  f_{\lambda}(\beta^g_c)\ N_{\tau}\,174\,{\rm MeV} = 5.07\,{\rm MeV}
\end{equation}
independently of the $N_{\tau}$ value.

For the 3D DLT we use different coupling constants in the two layers.
Along the lines of the usual derivations \cite{BB07} the temperature
in the entire DLT is given by $T=1/[a(\beta^g_x)\,N_{\tau}]$, $x={\rm 
in, \, out}$ (\ref{eq:T}). We just have a system for which the coupling 
depends on the space position as it does in other systems like, for 
instance, spin glasses.  The physical temperature does vary because 
the lattice spacing depends on the coupling $\beta^g=6/g^2$. When using 
two different $\beta^g$ values, the lower one gives a lower physical 
temperature. We choose it to correspond to a temperature in the 
confined phase and call its layer ``outside''. The $\beta^g$ value 
in the other layer, called ``inside'', will be iterated to find the 
maximum of the Polyakov loop susceptibility for its layer, which is 
our signal for the deconfining transition. To each plaquette we assign 
one of these values, $\beta^g_{\rm in}$ and $\beta^g_{\rm out}$, in a 
slightly asymmetrical way: If any link of plaquette is from the
inside layer, $\beta^g_{\rm in}$ is used, otherwise $\beta^g_{\rm out}$.
So the inside lattice becomes slightly larger than the outside one.

\section{Numerical results \label{sec:results} }

\begin{table} \begin{center} \scalebox{1.0}{%
\begin{tabular}{c||c|c|c||c|c|c|}
\textbf{$T_{out}$} & \multicolumn{3}{c}{158.15 MeV} & 
\multicolumn{3}{c}{161.80 MeV} \\ \hline \hline
\textbf{$N_{\tau}$} & \textbf{4} & \textbf{6} & \textbf{8} 
& \textbf{4} & \textbf{6} & \textbf{8} \\ \hline \hline
\textbf{$\beta^g_{out}$} &5.6500 &5.84318 &6.00458 & 5.6600 &5.85514 &
6.01821 \\ \hline \hline
$N_s$ & 20 & 30 & 40 & 20 & 30 & 40 \\ \hline 
$N_s$ & 24 & 36 & 48 & 24 & 36 & 48 \\ \hline 
$N_s$ & 28 & 42 & 56 & 28 & 42 & 56 \\ \hline 
$N_s$ & 32 & 48 & 64 & 32 & 48 & 64 \\ \hline 
$N_s$ & 36 & 54 & - & 36 & 54 & - \\ \hline 
$N_s$ & 40 & 60 & - & 40 & 60 & - \\ \hline 
\end{tabular}} 
\caption{Overview of our MCMC runs. \label{tab:run}}
\end{center} \end{table}

\begin{figure} \begin{center}
\includegraphics[width=0.7\textwidth]{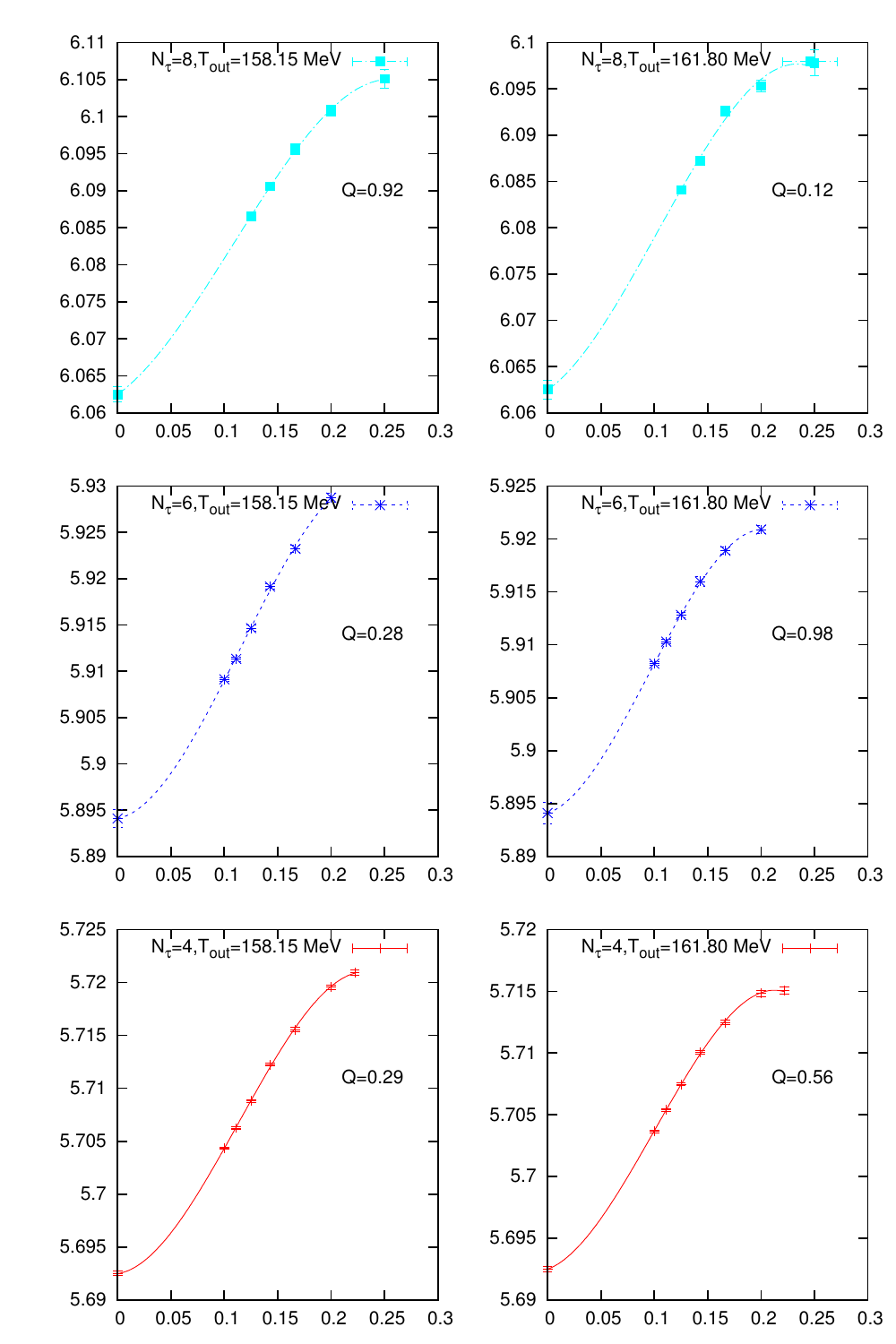} 
\caption{Pseudo transition couplings from maxima of the
         Polyakov loop susceptibility.} \label{fig_bfits}
\end{center} \end{figure} 

In our simulations we accommodate both layers in the SU(3) scaling 
region, so that (\ref{eq:LambdaL}) holds, which starts slightly below 
$\beta^g=5.65$. With our smallest $N_{\tau}=4$ temporal extension this 
sets the lower limit on $\beta^g_{\rm out}$. To work with round numbers
we decided to start $N_{\tau}=4$ simulations at $\beta^g_{\rm out}=5.65$ 
and $\beta^g_{\rm out}=5.66$. Translated into temperatures this 
corresponds to $T_{\rm out}=158.15\,{\rm MeV}$ and $T_{\rm out}=
161.80\,{\rm MeV}$. For larger $N_{\tau}=6$ and $N_{\tau}=8$ lattice 
these numbers have to be translated back into coupling constant values, 
listed in table~\ref{tab:run} where we give an overview of the 
performed runs. The accuracy of the given numerical precision of 
the $T_{\rm out}$ values reflects the SU(3) scaling relation when
translating back and forth, but has no significance for the accuracy 
of the physical temperature.

\begin{figure} \begin{center}
\includegraphics[width=0.49\textwidth]{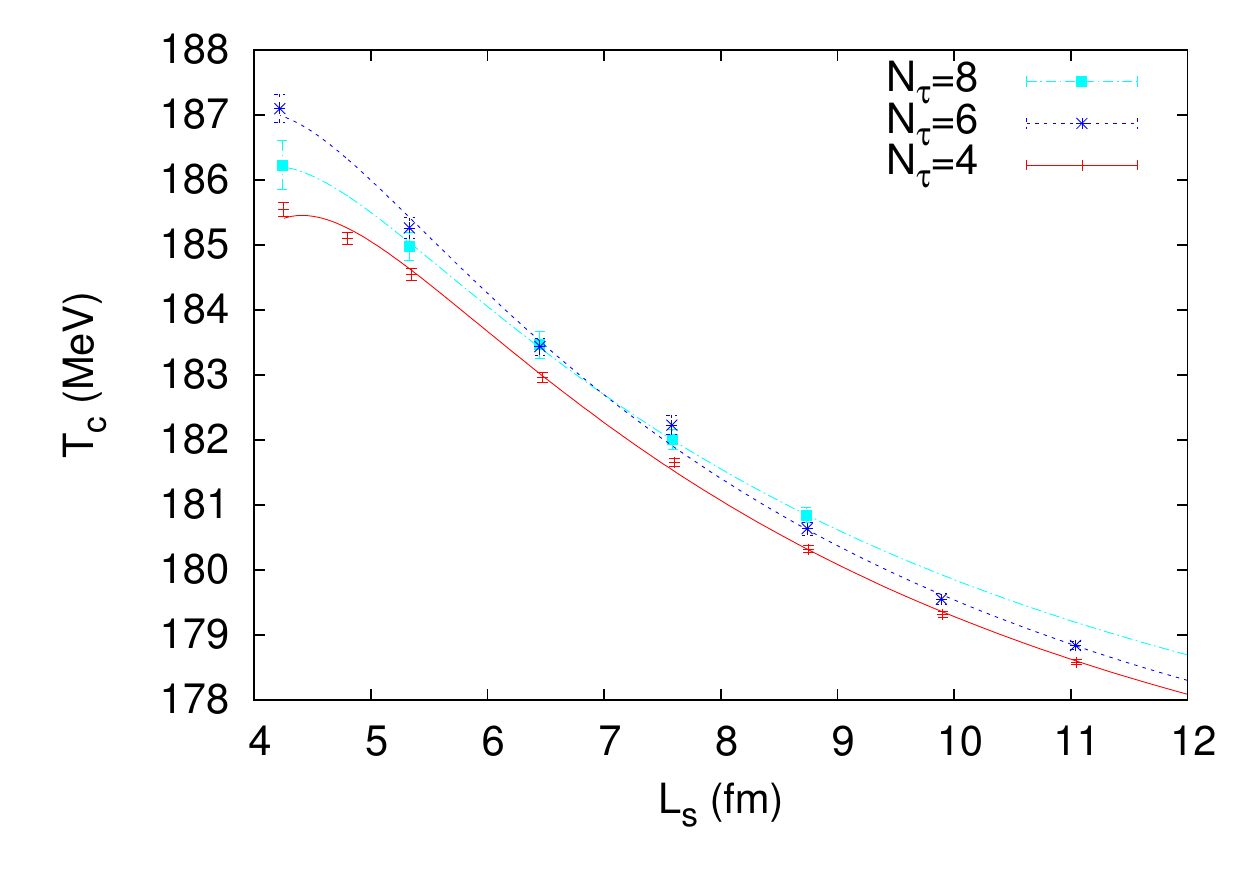} 
\includegraphics[width=0.49\textwidth]{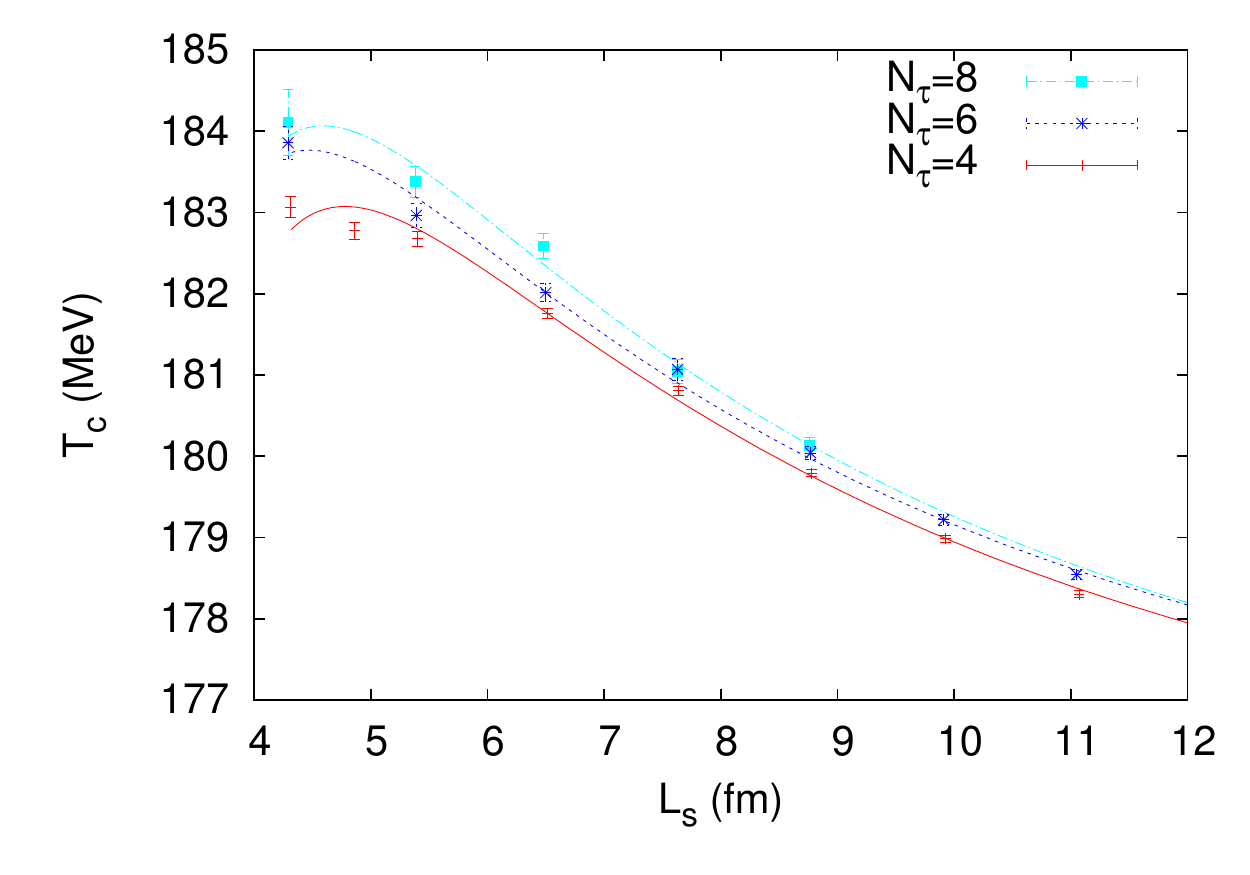} 
\caption{Pseudo transition temperatures versus volume edge length.}
\label{fig_Tcfits}
\end{center} \end{figure} 

The small temperature difference between the confined phase 
and the infinite volume transition temperature makes it rather CPU 
time demanding to locate our deconfinment signals, the maxima of the 
Polyakov loop susceptibility. This is in essence overcome by brute 
force computer time. Relying on Fortran MPI code which is documented 
in \cite{BW09}, the bulk of our production was run at NERSC on their
Cray XT4 and (more recently) Cray XE6, while code development and 
very limited production was done at FSU. All our runs used a 
statistics of $2^{13}$ sweeps for equilibration and $2^{21}$
sweeps for data production to overcome the statistical noise.
Improved measurements techniques were tried, but did not result in 
major CPU time gains. Production on a $4\times 32^3$ DLT used 1,024 
processors and took 5.7~hrs in real time. Approximate run times on 
other lattices can be scaled from this. 

\begin{figure} \begin{center}
\includegraphics[width=0.8\textwidth]{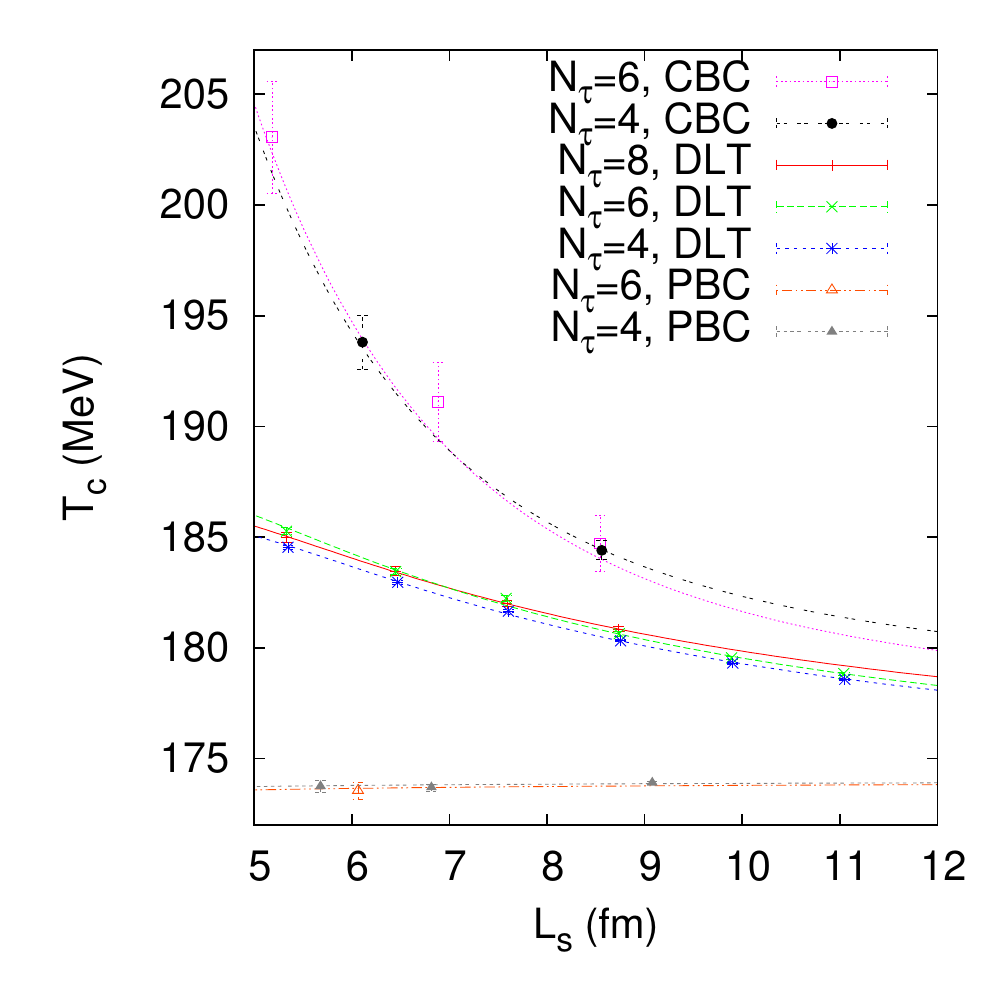} 
\caption{Pseudo transition temperatures versus volume edge length for 
CBC, DLT and PBC} \label{fig_TcLs}
\end{center} \end{figure} 

Fig.~\ref{fig_bfits} plots the raw data for pseudo transition couplings 
$\beta^g_c$ as determined from maxima of the Polyakov loop susceptibility 
and plotted versus the lattice ratio $N_{\tau}/N_s$ together with
polynomial fits in $N_{\tau}/N_s$ up to order $(N_{\tau}/N_s)^3$.
For $N_{\tau}=4$ and $N_{\tau}=6$ we have excluded data from our 
smallest lattices, $N_{\tau}/N_s \le 0.5$, because they suffer from 
peculiar finite size effects, which can be traced to the edges of 
the DLT layers. From down to up the lattices sizes are given by 
$N_{\tau}=4,\,6,\,8$. The corresponding $\beta^g$ ranges of pseudo 
transition temperatures shown in the figure increase from [5.69,5.725] 
($N_{\tau}=4$) to [5.89,5.93] ($N_{\tau}=6$) to [6.06,6.11] 
($N_{\tau}=8$).
As seen in Fig.~\ref{fig_Tcfits} the curves from these disconnected
regions collapse into almost one curve when the SU(3) scaling relation 
(\ref{eq:LambdaL}) is applied to convert $\beta^g_c$ to $T_c$ in units
of MeV, which we plot versus the volume edge length in fermi.

Our final Fig.~\ref{fig_TcLs} presents $T_c$ results for 
outside temperature $T_{\rm out} = 158.15\,{\rm MeV}$ together with 
those of CBC from \cite{BB07}, where the strong coupling limit $\beta^g
\to 0$ was used as BC. Even with our present high outside temperature 
the correction to the infinite volume critical temperature turns out to 
be remarkably large. So, this effect warrants further investigation.
Instead of using a DLT one may incorporate cold boundaries by a 
surrounding region and, in particular, one may be interested in 
the inclusion of quarks in such calculations.

\section{Summary and conclusions \label{sec:summary} }

\begin{itemize}

\item Even with a large outside temperature of about 160 MeV finite size 
correction of the SU(3) deconfining temperature due to small volumes 
are clearly visible.

\item 
Our estimates of (pseudo) transition temperatures show SU(3) scaling 
behavior when increasing the temperature extension of the lattice 
from $N_{\tau}=4$ to $N_{\tau}=6$ and $N_{\tau}=8$.

\item 
Small volumes increase the SU(3) transition temperature, while quark 
effects decrease it. If the increase holds also with quarks included, 
previous lattice calculations would underestimate the transition 
temperatures at RHIC.

\end{itemize}
\medskip

{\bf Acknowledgments:} We thank Alexei Bazavov for useful discussions. 
This work has in part been supported by DOE grant DE-FG02-97ER-41022. 
Most MCMC data were produced at NERSC under grant ERCAP 84105.

\end{document}